# Glycine's Radiolytic Destruction in Ices:
# First *In-Situ* Laboratory Measurements for Mars

Perry A. Gerakines and Reggie L. Hudson

Astrochemistry Laboratory, NASA Goddard Space Flight Center, Greenbelt, MD 20771

**Abstract:**

We report new laboratory studies of the radiation-induced destruction of glycine-containing ices for a range of temperatures and compositions that allow extrapolation to Martian conditions. *In-situ* infrared spectroscopy was used to study glycine decay rates as a function of temperature (from 15 to 280 K) and initial glycine concentrations in six mixtures whose compositions ranged from dry glycine to $H_2O$ + glycine (300:1). Results are presented in several systems of units, with cautions concerning their use. The half-life of glycine under the surface of Mars is estimated as an extrapolation of this data set to Martian conditions, and trends in decay rates are described as are applications to Mars's near-surface chemistry.

## 1. Introduction

The formation and survival of molecules in extraterrestrial environments is a subject of considerable interest to the astronomical community. In particular, it would be valuable from the standpoint of astrobiology to know how organic molecules withstand the conditions of high radiation flux and low temperature found in the surface layers of many planets and satellites. Cold planetary near-surface environments could be reservoirs of carbonaceous materials from past impact events or could hold the chemical evidence of past biological systems. It is possible that chemical tracers might decay over the course of time when exposed to radiation, and so it is important to determine the radiolytic lifetimes of these molecules when planning missions to other worlds. Understanding the survivability of biological molecules such as amino acids also is valuable to the field of planetary protection, where this information may aid in the determination of the levels of cross-contamination of terrestrial and potential extraterrestrial biospheres.

Despite the fact that spacecraft have orbited, landed on, and sampled Mars since the 1960s, large gaps remain in what we know about molecular evolution under Martian conditions. For instance, the fate of organic molecules delivered by meteorite impacts and their interactions with the oxidants in Martian soil has been the subject of speculation for years (Benner *et al.*, 2000). The relatively high costs of Mars missions have meant that heavy reliance is placed on laboratory measurements and theoretical models to unravel the thermal (Marion *et al.*, 2010), photo- (Stalport *et al.*, 2009), and radiation (Kminek and Bada, 2006) chemistries of suspected

Martian molecules. It remains a challenge to design such measurements and models so that they are, to the extent possible, realistic replicas of Martian processes and conditions.

Among the external drivers of Martian chemistry are photons and particle radiation, both readily penetrating the Martian $CO_2$ atmosphere (Molina-Cuberos *et al.*, 2001). The development and application of models of Martian chemistry has been hindered by the fact that the radiation environment on Mars has not been measured directly, with the exception of preliminary results from the MSL/Curiosity rover (Hassler *et al.*, 2012). Despite the dearth of radiation data from the surface of the planet, the three processes shown in Figure 1 appear to be agreed upon as the sources of energy for radiation chemistry on Mars at various depths (Dartnell *et al.*, 2007). Molecules on or within micrometer-to-millimeter of the Martian surface are thought to be altered significantly by far-UV solar photons (Stoker *et al.* 1997; ten Kate *et al.*, 2005). Starting at a few millimeters, extending to a few centimeters, and to perhaps a meter, solar energetic particles are the dominant source of ionizing radiation (Kminek and Bada, 2006). From at least a meter to several meters it is galactic cosmic rays that dominate, but their penetration and that of the secondary electrons they produce falls off with depth. By about 4 – 5 meters the greatest source of radiation in the Martian subsurface is from radioactive materials in the soil (Dartnell *et al.*, 2007).

The centimeter-to-meter depths below the Martian surface are where conventional radiation chemistry controls molecular change. Theoretical models have been published to aid in understanding molecular destruction at these depths (Dartnell *et al.*, 2007), but in practice these usually are based on room-temperature laboratory data, studies of single-component samples, and *ex situ* methods of chemical analysis. While such models and laboratory investigations contribute to a "baseline understanding" of sub-surface radiation chemistry on Mars, what also are needed are experiments on solid-phase mixtures at lower temperatures, and coupled with *in situ* analytical methods. Somewhat remarkably, few such experiments have been published in the 40 years since the Viking landers' Martian investigations, or earlier.

Our research group has used IR spectroscopy and ionizing radiation to study the formation and destruction of organic molecules in ices for many years. In our most-recent paper we examined the radiolytic destruction of three amino acids at 15 – 140 K in the presence and the absence of $H_2O$-ice (Gerakines *et al.*, 2012). Decay parameters were determined for only two mixtures (dry glycine and $H_2O$ + gly 8.7:1) and lifetimes were estimated for various astronomical environments based on extrapolations of this limited data set. In the present paper we report new and more extensive *in situ* laboratory measurements of the rate of radiation-induced destruction of the simplest amino acid, glycine ($H_2N-CH_2-COOH$, abbreviated "gly") over a larger range of temperatures (15 to 280 K) and concentrations (0:1 to 300:1). Our aim is to report these new data that allow extrapolation to Martian conditions and to illustrate the caution needed in describing radiolytic stabilities.

## 2. Experimental Methods

The methods used were similar, and in most cases identical, to those of our previous work (Gerakines *et al.*, 2012). Briefly, reagent-grade glycine was sublimed from a Knudsen-type furnace onto a polished aluminum substrate (area ≈ 5 cm$^2$) held at the desired temperature within a vacuum chamber (∼ 10$^{-7}$ torr) interfaced to an infrared (IR) spectrometer. Spectra of ices on the substrate were recorded as 100-scan accumulations at 4 cm$^{-1}$ resolution from 5000 to 700 cm$^{-1}$ using a Thermo Nicolet Nexus 670 spectrometer. A closed-cycle helium cryostat (ARS DE-204) and a resistive heater served to maintain the glycine-containing ice at the desired temperature within the 15 – 300 K range studied. In most experiments, H$_2$O vapor was admitted into the vacuum chamber to condense simultaneously with glycine onto the pre-cooled substrate. By careful choices of flow rates, H$_2$O:glycine ratios from 300:1 down to 0:1 could be attained. A typical H$_2$O sample growth rate was 5-10 μm hr$^{-1}$ The background pressure inside the vacuum chamber was typically ∼ 10$^{-7}$ torr, resulting in a residual H$_2$O condensation rate of ∼ 0.05 μm hr$^{-1}$. Ice irradiations were performed with 0.8 MeV protons from a Van de Graaff accelerator, a beam current of ∼ 0.1 μA, and a fluence of $1 \times 10^{14}$ p+ cm$^{-2}$ delivered in about 15 minutes. Direct measurement of the beam current in the underlying aluminum substrate allowed radiation doses to be calculated using reasonable assumptions for each sample's density (Weast *et al.*, 1984) and stopping power (Ziegler *et al.*, 2010). See Table 1 for the values used in this work. Sample thicknesses were on the order of 1-4 μm, measured by 650-nm laser interferometry. See Gerakines *et al.* (2012) for additional experimental details related to amino-acid sublimation, characterization, and irradiation.

It should be emphasized that the design of our experiments allowed amino-acid decay data to be collected *in situ* at the temperature of each irradiation. Raising the samples to room temperature for chemical analyses, as is commonly the case, was therefore not necessary. Also, the sublimation and co-condensation method we selected for preparing H$_2$O + glycine ices avoided the uncertainties in freezing room-temperature aqueous glycine solutions, which will crystallize to give regions of varying glycine concentration in the sample. In short, the value of our results is enhanced both by the way the samples were prepared and by the method of analysis.

## 3. Results

This paper presents results for solids containing a single amino acid, glycine. Aside from its simplicity, glycine was selected as it is the only amino acid found both in a comet's tail and in meteorites (Elsila *et al.*, 2009; Degens and Bajor, 1962). Also, our earlier work (Gerakines *et al.*, 2012) showed that glycine's rate of radiolytic destruction in H$_2$O mixtures is somewhat greater than for more-complex amino acids, so that glycine's behavior could be taken as a limiting case.

Our choice of H$_2$O-ice was suggested by its possible subsurface presence on Mars and its existence at the Martian polar caps. The high-temperature limit of the experiments we report here with H$_2$O + glycine ices was determined by H$_2$O-ice's rapid sublimation above about 160 K and its tendency to crystallize above about 155 K, two effects we wished to avoid.  The other prominent Martian ice, frozen $CO_2$, has a more-limited range of temperatures for stability in our vacuum system.  Nevertheless, future experiments are envisioned with amino acids in solid $CO_2$.

Glycine has three common polymorphs (crystalline forms), but the broad IR features of our samples, both with and without $H_2O$-ice, suggested that all of our ices were amorphous.  The molecule also has two structures, shown in Figure 2.  Condensation of pure glycine at 140 K gave the zwitterion form of the molecule, while condensation near 15 K gave the non-zwitterion structure.  These forms were confirmed in our samples by comparing our measured IR spectra to those found in the literature (e.g., Maté *et al.*, 2011).  Warming to about 100 K of any sample of pure glycine deposited near 15 K gave a rapid conversion to the zwitterionic form of the molecule, and samples retained their zwitterion structure upon re-cooling.  All ices to be irradiated were deposited at 140 K in order to create the zwitterionic structure for all experiments, as it is what is expected at Martian temperatures.

Representative IR spectra of irradiated samples are given in Figures 3 and 4.  Figure 3 shows spectra of glycine at 15 K before and after proton irradiation to a final fluence of $8 \times 10^{14}$ p+ $cm^{-2}$, while Figure 4 shows the spectra from a similar experiment for a $H_2O$ + glycine (50:1) mixture at 15 K.  Doses are given in the figures in units of eV molecule$^{-1}$.  See Gerakines *et al.* (2012) and references therein for spectral assignments in the IR.

By integrating the glycine $COO^-$ vibrational feature at 1408 $cm^{-1}$ before and after successive irradiations, data were collected for the decay curves shown in Figures 5 and 6.  In both figures, the left-hand panel displays results for four $H_2O$:gly mixture ratios (0:1, 8.7:1, 50:1, and 300:1) irradiated at 15 K, while the right-hand panel shows the data for the 50:1 mixture irradiated at 15, 100, and 140 K.  In Figure 5 the horizontal scale is energy absorbed per gram of sample, found by multiplying proton fluence ($F$, in p+ $cm^{-2}$) and stopping power ($S$, in eV $cm^2$ $g^{-1}$ $p+^{-1}$) for each ice.  The radiation dose is given in megagray units along the upper horizontal scale, where 1 MGy = $6.242 \times 10^{21}$ eV $g^{-1}$.  Doses in the older, but still common, unit megarad (Mrad) can be obtained from the relation 1 MGy = 100 Mrad.  Although doses in the radiation-chemical literature commonly are given in eV $g^{-1}$, MGy, and Mrad units, it is not unusual to find doses in the *astro*chemical literature in units of eV molecule$^{-1}$, calculated from the product $m_{avg}SF$ where $m_{avg}$ is the average mass (in g) of the ice sample's molecules.  Since $m_{avg}$ varies among the ices we studied, the relative positions of decay curves on an eV molecule$^{-1}$ scale will differ from those on an eV $g^{-1}$ (or Mrad or MGy) scale as can be seen by comparing the curves of Figures 5 and 6.  For example, the decay shown in the $H_2O$ + gly (8.7:1) ice sample at 15 K (left-hand panels) is similar to that of the 0:1 sample in Figure 5, but in Figure 6 the differing horizontal scales, determined by two very different values of $m_{avg}$, causes the 8.7:1 sample to display a much steeper slope.  However straightforward the relationship between the two horizontal scales may seem, it is important to point out differences due to inconsistencies in the way that relative decay rates are reported in the literature and to stress that the units do matter.  They are determined in units equivalent to eV $g^{-1}$ by some (*e.g.*, Baratta *et al.*, 2002 or Kminek and Bada, 2006) and in eV molecule$^{-1}$ by others (*e.g.*, Gerakines *et al.*, 2000 or Ferrante *et al.*, 2008), but the two do not necessarily reveal the same relative trends.  Tables 2 and 3 give half-life doses measured in our experiments, denoted $D_{1/2}$ and defined as the dose required for the destruction of one half of each sample's glycine molecules.

The initial density, sample area, and thickness of each ice we investigated could be measured directly by us or approximated from the mixing ratio and the known values for $H_2O$ and glycine. From this we calculated the number of glycine molecules present after the deposition of each sample, and graphs of the number of glycine molecules versus total energy absorbed, such as shown in Figure 7, were constructed. The initial slopes of the curves shown lead to glycine's radiation-chemical destruction rate, $G$(-gly), defined as the number of glycine molecules destroyed per 100 eV absorbed by the sample. This gives another trend, consistent with the manner in which decay rates are typically reported. The $G$(-gly) values we obtained by this procedure are given in Table 4 for each ice investigated.

## 4. Discussion

*4.1. Radiation chemistry of $H_2O$ + glycine mixtures*

Although this paper is not about the reaction products of ices containing glycine, we note that $CO_2$ formation was readily seen in all of our experiments. For example, both Figures 3 and 4 show an increase in the $CO_2$ absorbance at 2343 cm$^{-1}$. The decarboxylation ($CO_2$ release) of irradiated amino acids has been reported by us (Gerakines *et al.*, 2012) and many others (*e.g.*, Bonifačić *et al.*, 1998). Under our conditions, $CO_2$ could be formed by molecular decomposition of excited glycine molecules, to make $CO_2$ and methylamine, or by more-complex pathways as described below.

Our previous paper on amino-acid destruction (Gerakines *et al.*, 2012) included reactions and mechanisms for amino-acid destruction, focusing on direct decomposition of glycine and other molecules. However, here our interest is primarily in glycine at low concentrations in $H_2O$-rich ices, and a different chemistry applies. The two species directing chemical change in the early stages of radiolysis of $H_2O$-rich ices are hydrated electrons ($e_{aq}^-$), which promote reduction, and hydroxyl radicals (•OH), which promote oxidation. The reduction that destroys glycine is

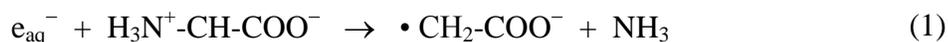
$$e_{aq}^- + H_3N^+\text{-CH-COO}^- \rightarrow \cdot CH_2\text{-COO}^- + NH_3 \qquad (1)$$

with the intense IR absorptions of $H_2O$-ice obscuring the infrared features of the ammonia produced. Hydroxyl radicals destroy the zwitterionic glycine in our ices through hydrogen-atom abstraction from the glycine's central carbon:

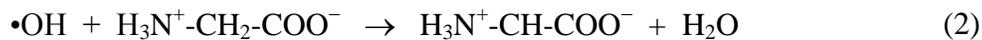
$$\cdot OH + H_3N^+\text{-CH}_2\text{-COO}^- \rightarrow H_3N^+\text{-CH-COO}^- + H_2O \qquad (2)$$

Disproportionation of the resulting zwitterionic radicals gives a molecule of glycine and a protonated imine ($H_2N^+=CH-COO^-$), the latter reacting with $H_2O$ molecules to produce ammonia and glyoxylic acid:

$$H_3N^+\text{-CH-COO}^- + H_3N^+\text{-CH-COO}^- \rightarrow H_3N^+\text{-CH}_2\text{-COO}^- + H_2N^+=CH\text{-COO}^- \quad (3)$$

$$H_2N^+=CH\text{-COO}^- + H_2O \rightarrow NH_3 + HC(=O)\text{-COOH} \quad (4)$$

It is possible that some of the glyoxylic acid produced is subsequently decomposed into carbon dioxide and formaldehyde ($H_2CO$), which would account for some of the $CO_2$ observed. See Neta *et al.* (1970) and Bonifačić *et al.* (1998) for additional information on these reactions.

*4.2. Radiation chemistry on Mars*

Although the multiple sources of ionizing radiation at Mars, and elsewhere, have been reviewed several times in this journal in recent years, much less attention has been paid to the differences between radiation chemistry, our area of study, and the related fields of nuclear chemistry and radiochemistry. Unlike nuclear chemistry the focus of radiation chemistry is neither transmutations nor radioactive decay, and unlike radiochemistry the focus is not the reactions of a single radioactive tracer. In radiation chemistry the emphasis is on the chemical changes induced by an incident ionizing agent. Specifically, as a keV-GeV ion (or electron or photon) passes through an icy solid it produces a trail of ionizations and excitations as the ion's energy is degraded. Many of these ionizations will produce secondary electrons which, in turn, travel through the ice creating separate tracks (δ rays) of yet more ionizations and excitations leading to even more chemical changes. In this way, which must be emphasized, the direct chemical action of the incident radiation is completely overshadowed by the chemistry produced by secondary electrons of energies less than about 20 eV (Pimblott and LaVerne, 2007). This means that to a robust first approximation the final products of various ionizing radiations (*e.g.*, $e^-$, $H^+$, $He^+$, x-rays, and γ-rays) are identical and, however counter-intuitive, strongly resemble the products of conventional photochemistry (Spinks and Woods, 1964). For comparisons of IR spectra of ices subjected to radiolytic and photolytic processing, see Hudson and Moore (2001), Hudson *et al.* (2001), and Gerakines *et al.* (2000), and the discussion of Hudson and Moore (2000). For comparisons from other laboratories, see, *e.g.*, Baratta *et al.* (2002).

Another point worth emphasizing concerns the particle energies available at Mars and the chemistry they induce. It is perhaps tempting to think that investigations of Martian-subsurface radiation chemistry must focus on the effects of particles with energies in the GeV and higher ranges, since they dominate the radiation environment at depths of about 1 m (Dartnell *et al.*, 2007), and not on the changes induced by MeV ions such as we have used. However, while GeV ions penetrate more deeply into an ice than MeV ions, the energy delivered by the former per unit distance (*i.e.*, their stopping power) is several orders of magnitude smaller than that of MeV ions of the same type. Combined with the fact that GeV cosmic rays are of much lower abundance than those of smaller energy, the conclusion is that the radiation chemistry on Mars induced by MeV ions dominates that of GeV ions.

The laboratory results we report here apply to depths of a few meters on Mars, where our proton source readily reproduces the radiation doses expected over long times. Specifically, the Martian dose rate estimated by Dartnell et al. (2007) is on the order of 0.1 Gy yr$^{-1}$, or approximately $2 \times 10^{-8}$ eV molecule$^{-1}$ yr$^{-1}$, at a depth of 1 m. This implies that a molecule under the surface of Mars for about 50 million years will absorb about 1 eV, a dose delivered to an ice by our Van De Graaff accelerator in 20-to-30 minutes. In short, our glycine experiments employ Martian subsurface radiation doses.

*4.3. Laboratory radiation results*

Several important points emerge from an examination of our data and calculations. From Figures 5 and 6 and Tables 2 and 3 one sees that glycine's rate of radiation-induced destruction depends on both temperature and the presence of $H_2O$-ice. This is a trend that is not always obvious in previous work. It suggests that measurements on glycine's radiation stability in the absence of $H_2O$ ice may not be directly applicable to Mars or other environments.

Comparison of the left-hand panels of Figures 5, 6, and 7 also leads to an important conclusion, namely that in any discussion of glycine's radiation stability in one chemical environment compared to another the units of dose must be specified carefully. In the present case, Figure 5 shows little difference in the decay rates of glycine and $H_2O$ + glycine ices on an eV per gram basis, but differences are obvious with Figure 6's eV per molecule scale. Figure 7, with a scale of total energy absorbed, shows yet a different trend in the initial decay slopes, one which appears to be the opposite of that in Figure 6. All three figures accurately portray the kinetics, but clearly the choice of *x*-axis units can alter the assessment of glycine's radiation stability determined from a half-life dose ($D_{1/2}$). In contrast, the right-hand panels of Figures 5, 6, and 7 show the decay curves for the same ice composition at three irradiation temperatures. In these cases, the trends are nearly identical, since the conversion from one horizontal scale to the next is primarily dependent on the ice composition.

Another conclusion, taken from Figure 7 and Table 4, is that $G$(-gly) depends on the initial relative abundance of glycine. This is not surprising since $G$(-M) is a kinetic parameter referring to the initial destruction of a molecule M. Although commonly termed a radiation-chemical yield, $G$(-M) says nothing about the final equilibrium abundance of a molecule. As a kinetic quantity it is not surprising that when the concentration of M is decreased then its loss rate (-dM/dt) falls, as seen in Figure 8 in moving to the right.

Our experiments also emphasize the importance of distinguishing between the so-called *direct* and *indirect* actions of ionizing radiation. In the former, energy reaches glycine molecules directly, but energy transfers and chemical change in the latter are caused by $H_2O$ molecules and their radiation products (*e.g.*, e$^-$, H, OH, $H_2O_2$). Significantly, Figure 8 shows that at greater dilutions, where indirect radiolytic action dominates in ices, a limiting value of $G$(-gly) is approached. For a dilution of 300:1, the limits are approximately 0.30, 0.09, and 0.06 at 15, 100, and 140 K, respectively . Other things being equal, glycine will have a greater resistance to radiation-induced change when it is present in an ice in lower abundances. Figure 8 also shows that the value of *G*(-gly) falls as the irradiation temperature increases from 15 to 140 K.

*4.4. Glycine on Mars*

Macromolecular carbon has been found in Martian meteorites (McDonald and Bada 1995; Glavin *et al*. 1999), but present conditions do not appear to favor the synthesis of organics on Mars. However, it is expected that amino acids, including glycine, are brought to Mars and other Solar System objects by meteorites and perhaps by comets too. Any surficial Martian glycine will be destroyed by photochemistry, but sub-surface material can persist until altered by agents such as cosmic radiation, electrical discharges, or geological processes. If surface material is transported downward faster than it can be destroyed photochemically then its chances of survival will increase due to the shielding provided by soil and ices.

The purpose of this paper has been to report new data that can be used to understand and predict glycine's radiolytic fate on Mars and other worlds. However, the application of laboratory data and theoretical models to Martian surface chemistry requires assumptions to be made about surface composition. Dartnell *et al.* (2007) have published dose-rate estimates for three such models, with their model for pure-$H_2O$ ice being the closest to our experiments. Their calculations show that over a million years, $H_2O$-ice at a depth of 1 meter receives a dose of 0.31 MGy, falling to about 0.17 MGy at 2 meters. Denser surface materials (*e.g.*, rock) further reduce these doses, and the rock-ice model of Dartnell *et al.* (2007) indeed shows doses after 1 million years of 0.08 MGy and 0.01 MGy for depths of 1 and 2 meters, respectively. In all cases background radioactivity is expected to dominate other radiation sources below a depth of about 4.5 meters (Dartnell *et al.*, 2007).

Our warmest, most-dilute ice in Figure 8, at 140 K (-133 °C), has $G$(-gly) $\approx$ 0.05, and our Table 2 and Table 3 give half-life doses for it that can be applied to Mars. Selecting a value of $D_{1/2}$ = 6.9 eV molecule$^{-1}$ (Table 3) gives a fractional destruction rate of 0.5 (50%) in 3.5-4.5 $\times$ $10^8$ years at a depth of 1 m. One also can extrapolate to higher temperatures the limiting *G* values we report for our most-dilute glycine-containing ice, giving $G$ = 0.043 at 210 K on Mars.

In Figure 9, we show the estimated half life of glycine as a function of depth in the Martian subsurface, where the half-life dose for an $H_2O$ + gly (300:1) mixture at 140 K has been applied, along with absorbed dose data for the rock-ice model by Dartnell *et al.* (2007). Note that doses for the rock-ice model are generally lower than that for pure $H_2O$-ice, but likely represent a more realistic representation of the Martian surface. The drill on the Mars Science Laboratory/Curiosity rover is capable of sampling the soil to a depth of about 5 cm, where the estimated half-life of glycine in ice is about 2-3 $\times$ $10^8$ yr, which is a relatively short timescale when compared to the $\sim$ 3.5-4 billion years since the era when Mars was likely the most habitable.

## 5. Summary and conclusions

- *G* values for glycine radiolytic destruction, based on an *in situ* method of analysis, have been presented, and variations with temperature and initial abundance have been explored.  As *G* values for glycine destruction are found to depend on both temperature and the presence of $H_2O$-ice, published room-temperature data on pure glycine must be used with caution in Mars applications, as must laboratory data not obtained by *in situ* methods.

- Radiation half-life doses in eV $g^{-1}$ and eV molecule$^{-1}$ have been determined for glycine embedded in an amorphous solid ($H_2O$-ice).  Statements about the stability of glycine to radiolysis need to be accompanied by careful attention to dose units.

- Any glycine, or other amino acid, present on Mars is expected to have a low abundance, and so will be partially shielded by its surroundings, be they made of soil or ice, from the direct effects of ionizing radiation.  Our data provide a laboratory basis for estimating glycine destruction times in the case of a $H_2O$-ice environment.

- The differences in the direct and indirect effects of ionizing radiation are important in any application of laboratory data to problems of solar system chemistry.  Moreover, since the existence of extraterrestrial crystalline amino acids is unlikely, due to these molecules' low abundance, attention must be paid in laboratory studies to the amorphous or crystalline nature of amino-acid samples.

**Acknowledgments**

Marla Moore and Jan-Luca Bell contributed some of the early measurements of glycine and $H_2O$ + glycine used in this study.  Mark Loeffler is recognized for substantial assistance in day-to-day operations of the equipment in our laboratory.  Zan Peeters constructed and tested the sublimation oven we used.  The authors wish to acknowledge support from the NASA Astrobiology Institute through a grant to the Goddard Center for Astrobiology.  The support of NASA's Exobiology Program also is gratefully acknowledged.  Finally, we thank Steve Brown, Tom Ward, and Eugene Gerashchenko, members of the Radiation Effects Facility at the NASA Goddard Space Flight Center, for operation of the proton accelerator.

**References**

Baratta, G. A., Leto, G., and Palumbo, M. E. (2002)  A comparison of ion irradiation and UV photolysis of $CH_4$ and $CH_3OH$.  *Astron. & Astrophys.* 384:343-349.
Benner, S. A., Devine, K. G., Matveeva, L. N., and Powell, D. H. (2000) The missing organic molecules on Mars.  *Proc. Natl. Acad. Sci.* 97: 2425-2430.
Bonifačić, M., Štefanić, I, Hug, G. L., Armstrong, D. A., and Klaus-Dieter, A. (1998)  Glycine decarboxylation: The free radical mechanism.  *J. Am. Chem. Soc.* 120:9930-9940.

Dartnell, L. R., Desorgher, L., Ward, J. M., and Coates, A. J. (2007) Modelling the surface and subsurface Martian radiation environment: Implications for astrobiology. *Geophys. Res. Letters* 34:L02207.

Degens, E. T., and Bajor, M. (1962) Amino acids and sugars in the Bruderheim and Murray meteorite. *Naturwissenschaften* 49:605-606.

Ferrante, R. F., Moore, M. H., Spiliotis, M. M., and Hudson, R. L. (2008) Formation of interstellar OCS: Radiation chemistry and IR spectra of precursor ices. *Astrophys. J.* 684:1210-1220.

Gerakines, P. A., Moore, M. H., and Hudson, R. L. (2000) Carbonic acid production in $H_2O$ + $CO_2$ ices: UV photolysis vs. proton bombardment. *Astron. & Astrophys.* 357:793-800.

Gerakines, P. A., Hudson, R. L., Moore, M. H., and Bell, J.-L. (2012) In-situ measurements of the radiation stability of amino acids at 15-140 K. *Icarus* 220:647-659.

Glavin, D. P., Bada, J. L., Brinton, K. L. F., and McDonald, G. D. (1999) Amino acids in the Martian meteorite Nakhla. *Proc. Natl. Acad. Sci.* 96:8835-8838.

Hassler, D. M., Zeitlin, C., Wimmer-Schweingruber, R. F., Böttcher, S., Martin, C., Andrews, J., Böhm, E., Brinza, D. E., Bullock, M. A., Burmeister, S., Ehresmann, B.,·Epperly, M., Grinspoon, D., Köhler, J., Kortmann, O.,Neal, K., Peterson, J., Posner, A., Rafkin, S., Seimetz, L., Smith, K. D., Tyler, Y., Weigle, G., Reitz, G., and Cucinotta, F. A. (2012) The Radiation Assessment Detector (RAD) investigation. *Space Sci. Rev.* 170: 503-558.

Hudson, R. L. and Moore, M. H. (2000) New experiments and interpretations concerning the $OCN^-$ band in interstellar ice analogues. *Astron. & Astrophys.* 357:787-792.

Hudson, R. L. and Moore, M. H. (2001) Radiation chemical alterations in solar system ices: an overview. *JGR-Planets* 106:33275-33284.

Hudson, R. L., Moore, M. H., and Gerakines, P. A. (2001). The formation of cyanate ion ($OCN^-$) in interstellar ice analogues. *Astrophys. J.* 550:1140-1150.

Kminek, G. and Bada, J. L. (2006) The effect of ionizing radiation on the preservation of amino acids on Mars. *Earth Planet. Sci. Lett.* 245:1-5.

Maté, B., Rodriguez-Lazcano, Y., Gálvez, Ó., Tanarro, I., and Escribano, R. (2011) An infrared study of solid glycine in environments of astrophysical relevance. *Phys. Chem. Chem. Phys.* 13:12268–12276.

McDonald, G. D., and Bada, J. L. (1995) A search for endogenous amino acids in the Martian meteorite EETA79001. Geochim. Cosmochim. Acta 59:1179-1184.

Molina-Cuberos, G. J., Stumptner, W., Lammer, H., and Komle, N. I. (2001). Cosmic ray and UV radiation models on the ancient martian surface. *Icarus* 154: 216-222.

Neta, P., Simic, M., and Hayon, E. (1970) Pulse radiolysis of aliphatic acids in aqueous solution. III. Simple amino acids. *J. Phys. Chem.* 74:1214-1220.

Pimblott, S. M. and LaVerne, J. A. (2007) Production of low-energy electrons by ionizing radiation. *Rad. Phys. Chem.* 76:1244-1247.

Spinks, J. W. T. and Woods, R. J. (1964) *An Introduction to Radiation Chemistry*. Wiley and Sons, New York, p. 70.

Stalport, F., Coll, P., Szopa, C., Cottin, H., and Raulin, F. (2009) Investigating the photostability of carboxylic acids exposed to Mars surface ultraviolet radiation conditions. *Astrobiology* 9:543-549.

Stoker, C. R. and Bullock, M. A. (1997) Organic degradation under simulated Martian conditions. *J. Geophys. Res.* 102:10881-10888.

ten Kate, I. L., Garry, J. R. C., Peteers, Z., Quinn, R., Foing, B., and Ehrenfreund, P. (2005) Amino acid photostability on the Martian surface. *Met. Planet. Sci.* 40:1185-1193.

Weast, R.C., Astle, M.J., and Beyer, W.H. (Eds.) (1984) *The CRC Handbook of Chemistry and Physics, 64th ed.*, CRC Press, Boca Raton.

Ziegler, J.F., Ziegler, M.D., and Biersack, J.P. (2010) SRIM – the stopping and range of ions in matter. *Nucl. Instrum. Methods Phys. Res. Sect. B* 268:1818–1823.

Table 1. Properties of 0.8 MeV protons and amino-acid ice samples.

| Ice's H$_2$O:glycine ratio | Thickness ($h$) μm | Stopping power ($S$) eV cm$^2$ g$^{-1}$ p+$^{-1}$ | Density ($\rho$) g cm$^{-3}$ | Average molecular mass ($m_{avg}$) g molecule$^{-1}$ |
|---|---|---|---|---|
| 0:1 | 0.9 | $2.78 \times 10^8$ | 1.61 | $1.25 \times 10^{-22}$ |
| 8.7:1 | 1.8 | $2.86 \times 10^8$ | 1.06 | $3.85 \times 10^{-23}$ |
| 50:1 | 3.1 | $2.88 \times 10^8$ | 1.01 | $3.17 \times 10^{-23}$ |
| 100:1 | 3.0 | $2.89 \times 10^8$ | 1.01 | $3.08 \times 10^{-23}$ |
| 200:1 | 3.0 | $2.89 \times 10^8$ | 1.00 | $3.04 \times 10^{-23}$ |
| 300:1 | 4.0 | $2.89 \times 10^8$ | 1.00 | $3.02 \times 10^{-23}$ |

Table 2. Half-life doses in eV g$^{-1}$ / $10^{23}$

| Ice's H$_2$O:glycine ratio | Approximate temperature of ice (K) | | | |
|---|---|---|---|---|
| | 15 | 100 | 140 | 280 |
| 0:1 | $1.01 \pm 0.12$ | $1.57 \pm 0.38$ | $1.56 \pm 0.45$ | $1.21 \pm 0.24$ |
| 8.7:1 | $1.07 \pm 0.22$ | $1.03 \pm 0.26$ | $1.13 \pm 0.13$ | |
| 50:1 | $0.78 \pm 0.16$ | $1.46 \pm 0.29$ | $1.60 \pm 0.25$ | |
| 100:1 | $0.56 \pm 0.17$ | $0.54 \pm 0.11$ | $3.1 \pm 5.5$ | |
| 200:1 | $0.93 \pm 0.73$ | | $1.26 \pm 0.69$ | |
| 300:1 | --- | $0.93 \pm 0.32$ | $2.3 \pm 1.5$ | |

Table 3. Half-life doses in eV molecule$^{-1}$.

| Ice's H$_2$O:glycine ratio | Approximate temperature of ice (K) | | | |
|---|---|---|---|---|
| | 15 | 100 | 140 | 280 |
| 0:1 | $12.6 \pm 1.4$ | $19.5 \pm 4.7$ | $19.4 \pm 5.6$ | $15.1 \pm 3.0$ |
| 8.7:1 | $4.24 \pm 0.90$ | $4.1 \pm 1.0$ | $4.47 \pm 0.51$ | |
| 50:1 | $2.47 \pm 0.52$ | $4.64 \pm 0.91$ | $5.06 \pm 0.78$ | |
| 100:1 | $1.72 \pm 0.52$ | $1.65 \pm 0.32$ | $10 \pm 17$ | |
| 200:1 | $2.8 \pm 2.2$ | | $3.8 \pm 2.1$ | |
| 300:1 | --- | $2.79 \pm 0.98$ | $6.9 \pm 4.6$ | |

Table 4. $G$(-gly) values.

| Ice's H$_2$O:glycine ratio | Approximate temperature of ice (K) | | | |
|---|---|---|---|---|
| | 15 | 100 | 140 | 280 |
| 0:1 | $5.78 \pm 0.54$ | $3.55 \pm 0.42$ | $3.32 \pm 0.59$ | $4.52 \pm 0.49$ |
| 8.7:1 | $1.88 \pm 0.28$ | $1.80 \pm 0.26$ | $1.67 \pm 0.10$ | |
| 50:1 | $0.716 \pm 0.092$ | $0.516 \pm 0.058$ | $0.338 \pm 0.032$ | |
| 100:1 | $0.524 \pm 0.118$ | $0.611 \pm 0.085$ | $0.163 \pm 0.085$ | |
| 200:1 | $0.371 \pm 0.073$ | | $0.139 \pm 0.053$ | |
| 300:1 | $0.304 \pm 0.028$ | $0.088 \pm 0.018$ | $0.058 \pm 0.015$ | |

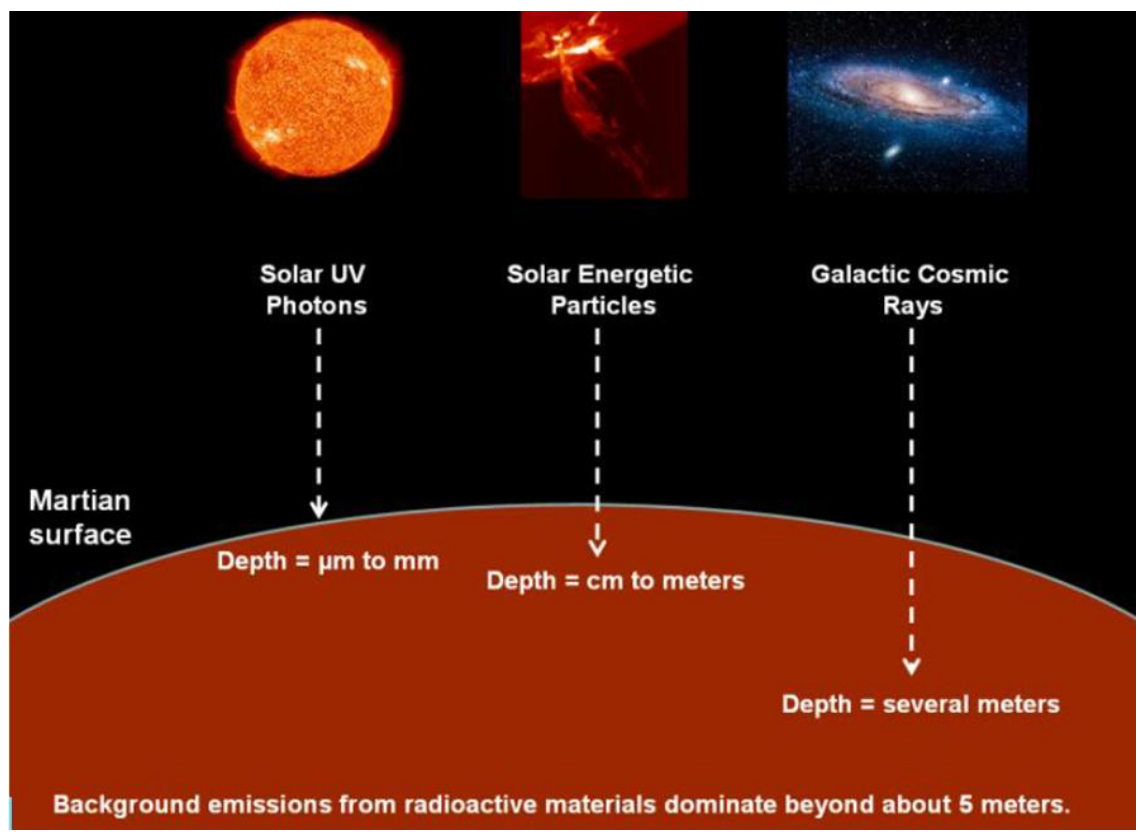

**FIG. 1.** Sources of radiation in the near-surface environment on Mars.

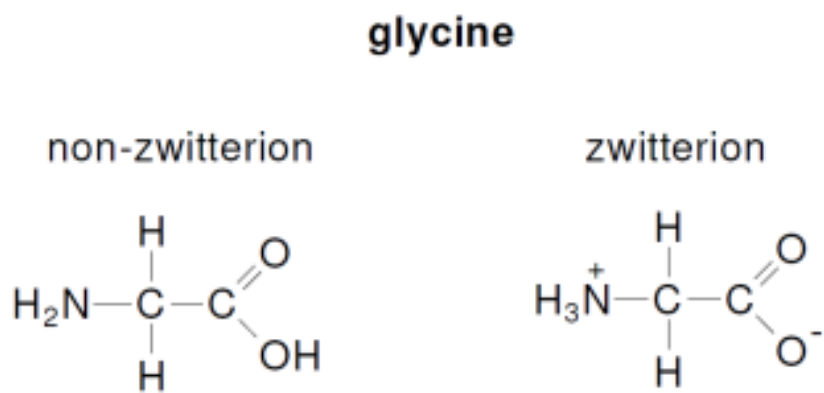

**FIG. 2.** Molecular structures of glycine.

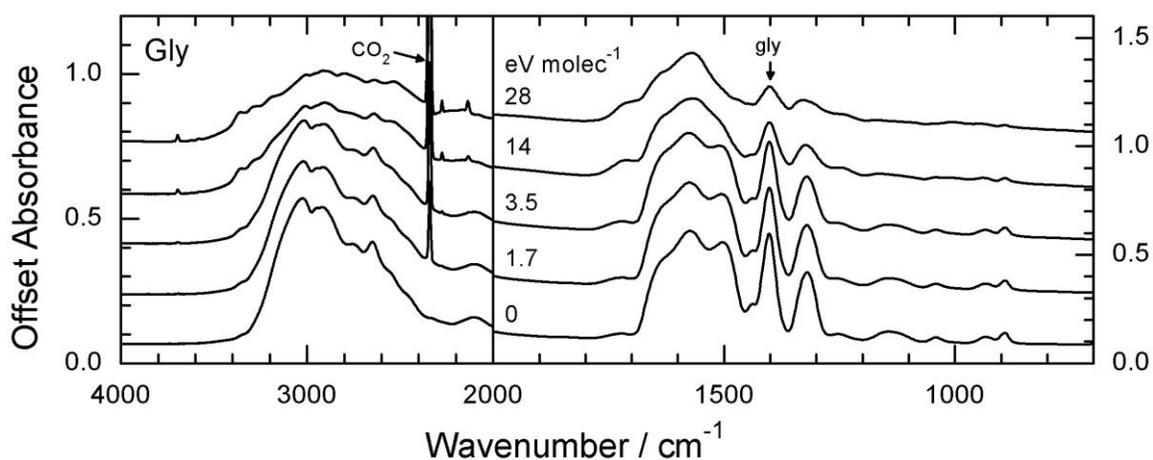

**FIG. 3.** IR spectra of glycine before and after irradiation at 15 K. The 2340 cm$^{-1}$ feature of the radiation product $CO_2$ and the 1408 cm$^{-1}$ feature of glycine are indicated.

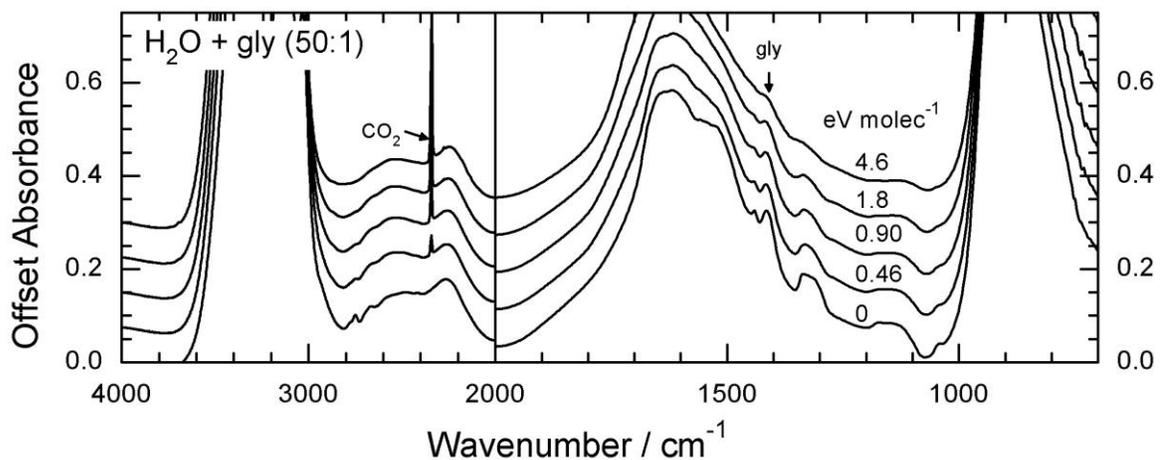

**FIG. 4.** IR spectra of $H_2O$ + gly (50:1) before and after irradiation at 15 K. The 2340 cm$^{-1}$ feature of the radiation product $CO_2$ and the 1408 cm$^{-1}$ feature of glycine are indicated.

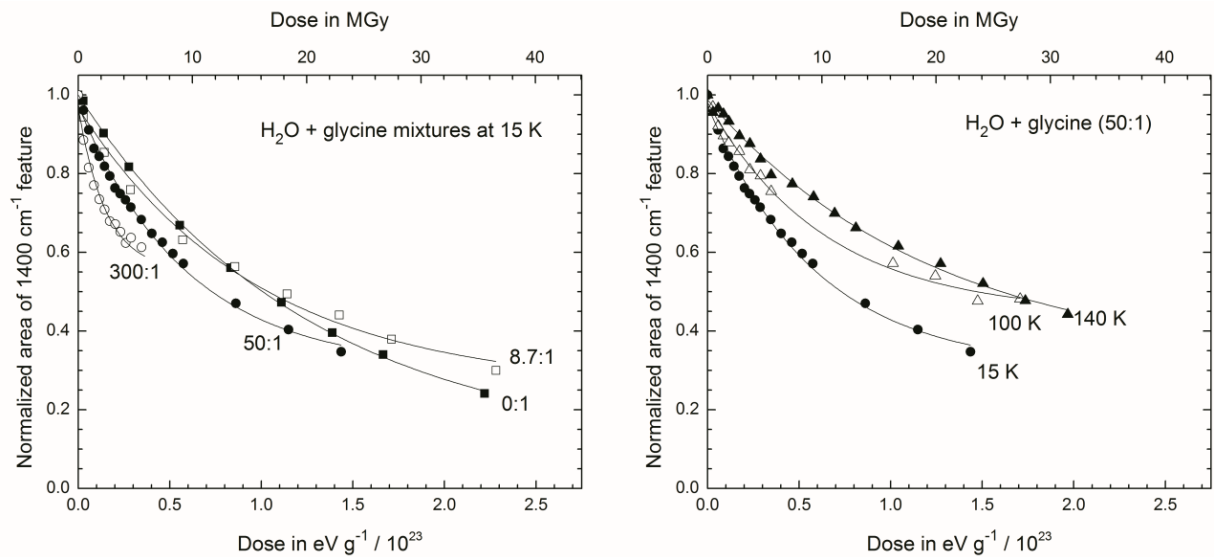

**FIG. 5.** Area of the glycine feature near 1400 cm$^{-1}$ during irradiation for selected ice compositions and irradiation temperatures, displayed as a function of dose in units of eV g$^{-1}$ (bottom axis) and MGy (top axis).

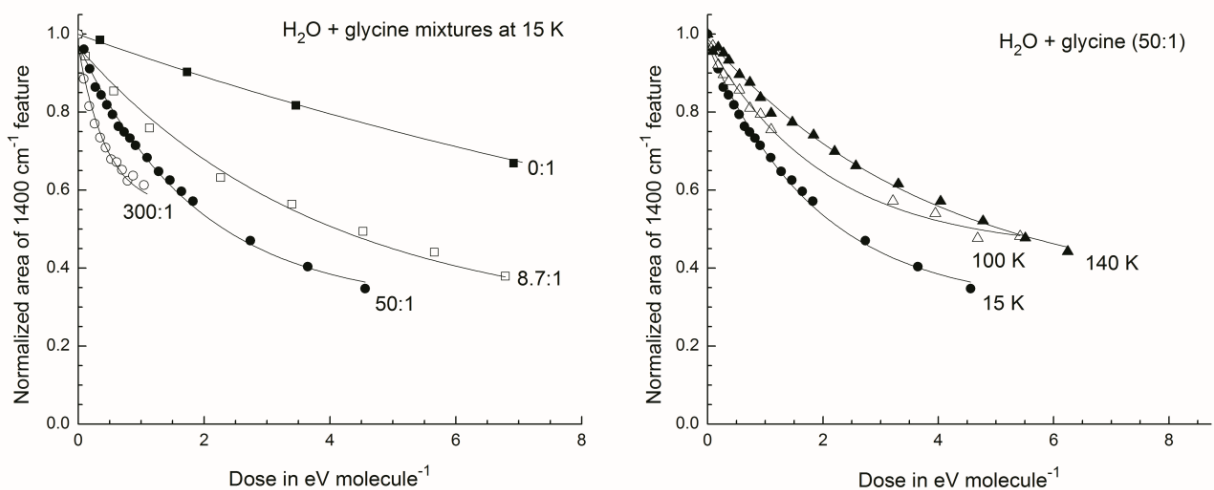

**FIG. 6.** Area of the glycine feature near 1400 cm$^{-1}$ during irradiation for the same selected ice compositions and irradiation temperatures shown in Figure 5, but displayed here as a function of dose in units of eV molecule$^{-1}$. In the left panel, data points for doses higher than 8 eV molecule$^{-1}$ were omitted for clarity.

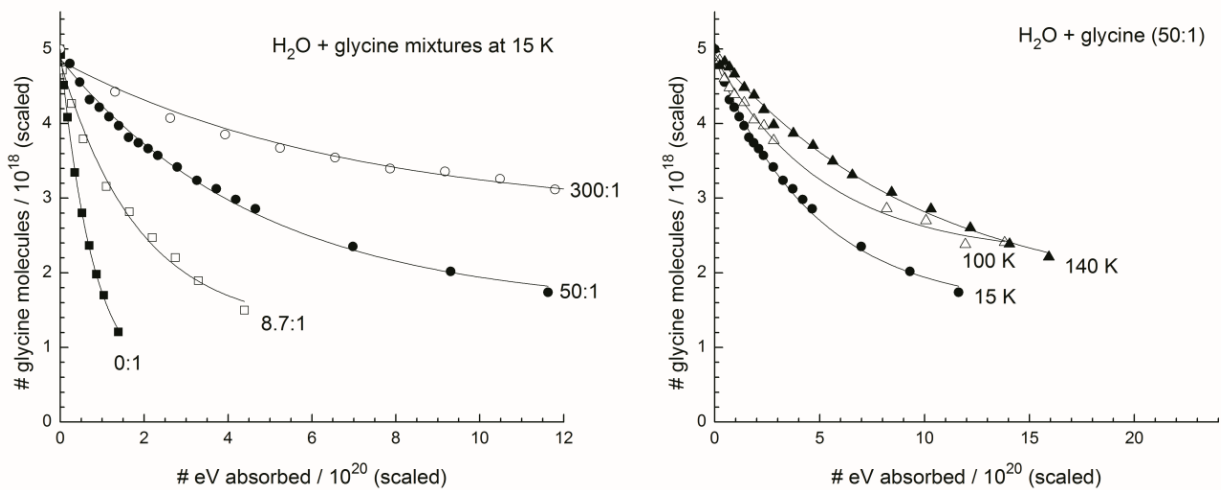

**FIG. 7.** The decay of glycine as a function of energy absorbed in units of eV for the same selected ice compositions and irradiation temperatures shown in Figures 5 and 6. Data for each mixture have been scaled to the same initial number of glycine molecules.

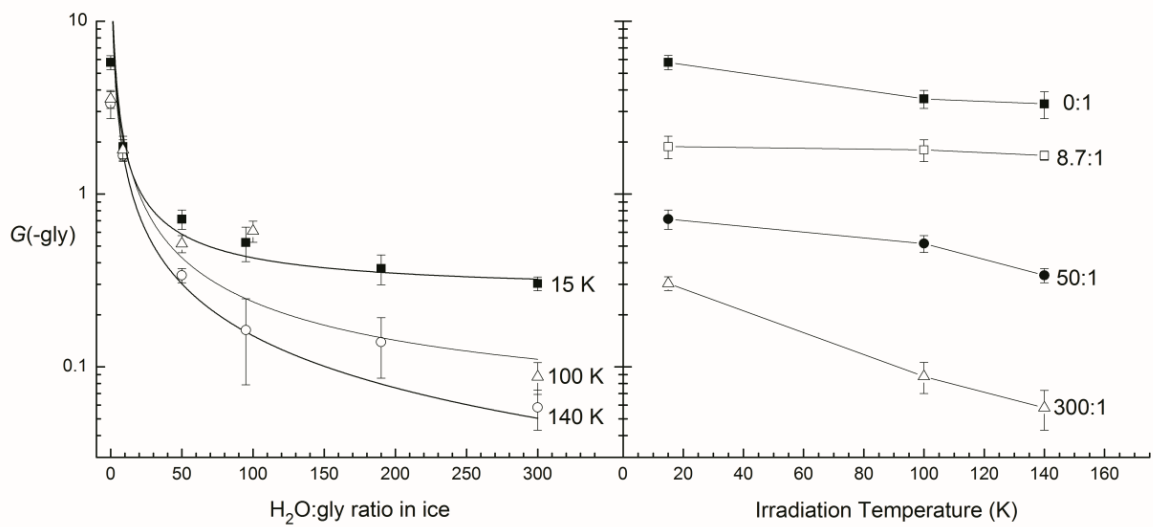

**FIG. 8.** $G$(-gly) as a function of ice composition (left) and irradiation temperature (right).

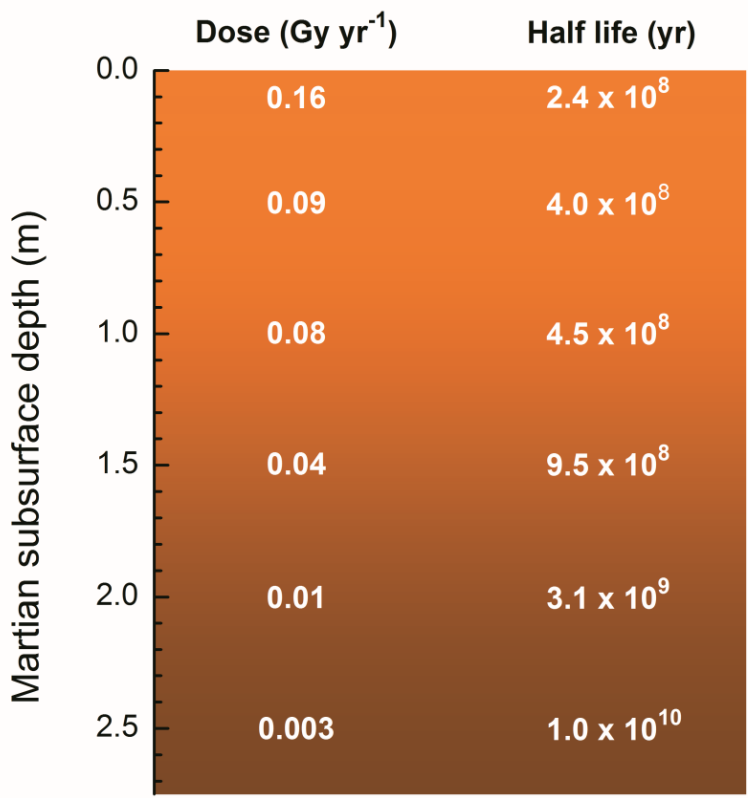

**FIG. 9.** The estimated half life for glycine mixed with $H_2O$-ice under the surface of Mars as a function of depth. Data are based on the half-life dose of glycine in our $H_2O$ + gly (100:1) mixture at 140 K (see Table 2) and the model of dose versus depth for an ice-rock mixture given by Dartnell *et al.* (2007).